# A simplified procedure to capture N($^4$S) kinetics during the transients of a strongly emissive pulsed ECR plasma using ns-TALIF

E. Bisceglia, S. Prasanna*, K. Gazeli, X. Aubert, C. Y. Duluard, G. Lombardi, K. Hassouni

LSPM, CNRS-UPR3407, Université Sorbonne Paris Nord, 99 Avenue J. B. Clément 93430 Villetaneuse.

*Corresponding Author

**Abstract**
A simplified and straightforward methodology that may be used to measure N($^4$S) atom density by means of nanosecond-two-photon absorption laser induced fluorescence (ns-TALIF) technique in strongly emissive plasmas is presented. The method makes use of the TALIF intensities measured with the laser central frequency tuned to the peak of absorption, instead of the fully integrated laser excitation spectrum. The use of this method, along with a physics-based fitting/filtering procedure enables a substantial reduction of the experimental uncertainty and allows performing measurement even in strongly emissive discharges. After an adequate validation, the method was employed to investigate the transients of a strongly emissive pulsed microwave discharge. We especially demonstrated the existence of an unexpected increase in N($^4$S) atom density at the early stage of the transition from high- to low-power phase. The use of a self-consistent quasi-homogenous plasma model taking into account a detailed state-to-state kinetics, enables to attribute this enhancement to surface de-excitation of N($^2$D) and N($^2$P) atoms.

Key-words : TALIF, nitrogen dissociation, microwave, pulsed plasma, kinetics.

## 1. Introduction

The monitoring of N-atom density in discharge plasmas and plasma post-discharges received a large attention from many research groups. This interest was motivated by the variety of applications that would benefit from the development of effective high-yield N-atom sources. These include material processing [1–3], biomedical applications [4,5] and Nitrogen-fixation [6,7].

The development of two-photon absorption laser induced fluorescence (TALIF) [8,9] brought many important improvements on the experimental determination of atomic densities in terms of space-and time-resolution, accuracy and flexibility. Two TALIF schemes were proposed for N-atom. The first one makes use of the [(2p$^3$) $^4$S$^0$, (3p) $^4$D$^0$, (3s) $^4$P] excited state system with a two-photon absorption at 2x211 nm and a subsequent fluorescence at 869 nm [8], while the second one makes use of the [(2p$^3$) $^4$S$^0$, (3p) $^4$S$^0$, (3s) $^4$P] system with a two photon absorption at 2x206.67 nm and subsequent fluorescence lines at 742.364, 744.230 and 746.831 nm [10]. The later system is generally preferred due to larger two-photon absorption cross section and more easily detectable fluorescence [11]. Whatever the adopted system, the use of TALIF to monitor absolute N-atom density requires calibration. While early calibration experiments were performed using NO-titration on dedicated setup [12], a straightforward and reliable calibration method based on the use of the TALIF signal of krypton was proposed two decades ago [13]. In fact, Krypton is a rare gas that provides a TALIF system, i.e., [(4p$^6$)$^1$S$_0$, 5p'[3/2]$_2$, 5s[3/2]$_1$], with wavelength values for laser-two-photon-absorption (2x204.13 nm) and fluorescence (587.09 nm) that are similar to those of N-atom. This method was successfully used for space- and time-resolved measurements of the absolute N-atom density in a variety of moderately-emissive plasmas and plasma post-discharges [14–25]. Surprisingly, despite this large success, achieving N-atom absolute density measurements in highly emissive nitrogen plasmas remains a challenging task from an experimental point of view [16,17,26]. As a matter of fact, N-atom fluorescence takes place inside the emission domain of the first positive system (FPS) of molecular nitrogen. It is therefore often very difficult to extract the fluorescence signal in strongly emissive non equilibrium nitrogen-containing plasma [16,26]. Several solutions have been proposed to circumvent this issue. The first type makes use of space and wave-length filtering around the probed volume and the fluorescence wavelength, respectively [11,16,18]. Unfortunately, although this approach usually results in the decrease in the FPS emission, it also leads to the decrease in the fluorescence intensity, which introduces significant uncertainty on the measured value of the N-atom density. The second approach used to get rid of the FPS emission consists of operating the plasma in a pulsed mode and performing TALIF measurements under post-discharge conditions during the plasma-off phase [18,26,27]. The N-atom absolute density in the plasma-on phase is estimated by time-extrapolating the absolute density-values determined during the post-discharge phase. In this case, it is usually assumed that N-atom depletion mechanism is driven by diffusion/wall recombination process [27]. Beside the uncertainty that may result from the extrapolation procedure, the validity of this approach requires the absence of fast kinetic effects that alter the N-atom density during the early stage of the plasma-off phase. For instance, the authors of reference [17] have clearly shown that the change in the discharge characteristics, electron density and temperature, etc., when transitioning from the plasma-on to the plasma-off phase in pulsed microwave plasma can yield unexpected N-atom time-variations during the very early post-discharge phase. In such a situation the determination of plasma-on N-atom density by extrapolating the time evolution of the atom density during the post-discharge phase is not valid. This shows that direct measurement of N-atom density during the plasma-on phase is still needed for such strongly emissive discharges.

This is one of the objectives of this article where we discuss time-resolved measurements of N-atom absolute density in low pressure high density pulsed ECR discharges that are excited with magnetized dipolar sources [28,29]. We propose and demonstrate the validity conditions of the simplified and approximate method where the atom density is inferred from the TALIF signal measured with the laser central frequency tuned to the peak of absorption, instead of the fully integrated laser excitation spectrum. We use this method along with optical emission spectroscopy (OES) measurements to investigate the processes that govern N-atom and $N_2(B)$ excited states kinetics during the transients in this pulsed ECR plasma.

It is worthy to mention at this stage that the use of the peak fluorescence signal to determine the N-atom-density has been already used in [16], where the authors investigated highly emissive discharge conditions for which the use of frequency-integrated laser excitation spectra is very difficult. Precise analysis, assessment and validation of the approach were however not considered in that reference.

The article is organized as follows. The experimental setup used in this study is presented in section 2. In section 3 we show how the methodology generally used to infer an absolute atom density from frequency integrated laser excitation spectra can be simplified provided that some constraints on the atomic absorption line broadening, that are likely to be fulfilled in cold non equilibrium plasma, are satisfied. We also discuss how the time and frequency profiles of the fluorescence signals are processed in order to properly filter out the strong background emission and determine the N-atom density during the plasma-on phase. The results are discussed in section 4 where the simplified method developed in this work is first validated by comparing the resulting N-atom density values with those obtained using frequency integrated laser excitation spectra. We then use this simplified approach along with OES to determine the N($^4$S)-atom density in this ECR plasmas as a function of the peak power and to investigate the processes that govern N-atom and N$_2$(B) excited states kinetics during the transients of the pulsed ECR plasmas under investigation. The interpretation that may be drawn from our measurements along with the comparison with other published works further validate the simplified approach proposed in this paper for non-equilibrium nitrogen containing plasmas that show moderate change in the gas temperature.

## 2. Experimental set-up

*2.1 The plasma source*

The microwave (MW) assisted plasma was generated using a commercial ECR source (Sairem Aura-Wave) detailed in [30]. The magnet is placed on the plasma source such that the magnetic field ensuring the ECR regime is directed toward the center of the reactor chamber, thus minimizing the electron losses at the walls. The magnetic field has an intensity of 0.0875 T and the resulting plasma has a toroidal shape (cf. figure 1). The MW power is supplied through a coaxial feed from a solid-state MW generator (Sairem GMS 200) that delivers a maximum power of 200 W. The operating MW frequency was tuned between 2.4 and 2.5 GHz in order to achieve optimal impedance matching between the source and the discharge system. The plasma may be operated in both continuous and pulsed modes. The pulsed mode regime makes use of square waveform power pulse and is characterized by the pulse period, the peak power and the duty cycle which represents the ratio of the high power phase duration to the pulse period. The source was embedded in a 100 mm-diameter stainless-steel 6-way cross and was placed on a specifically designed flange at the top of the cross. Three flanges were dedicated to optical viewports. The laser beam enters and exits the plasma reactor through two fused silica viewports (90% transmission at 205 nm). The fluorescence was collected perpendicularly to the laser beam through a borosilicate viewport. A metallic grid was placed at each viewport in order to prevent MW leakage outside the cavity. The vacuum was obtained using a set of turbo-molecular and rotary pumps (Edwards EXT75DX and Pfeiffer Duo 6M), allowing to achieve a residual pressure of 10$^{-5}$ Pa. The working pressure in the chamber was monitored using a capacitive gauge (Pfeiffer CCR363) working in the range 0.133-1333 Pa. The gases were injected in the reactor using mass flow controllers (Bronkhorst EL flow), the flowrate being set at 5 sccm.

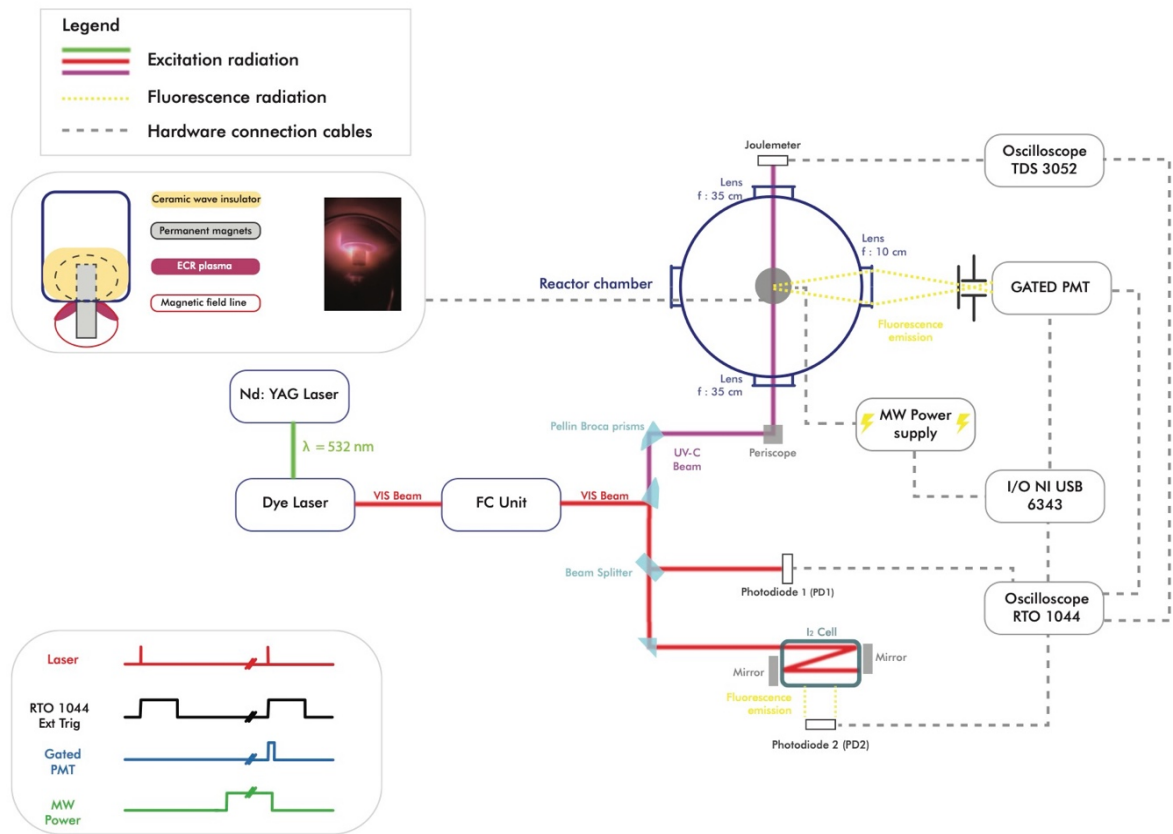

Figure 1: experimental setup

*2.2 The ns-TALIF setup*

The characterization of the pulsed MW plasma by means of a ns-TALIF diagnostic was carried out using a dye laser pumped by a pulsed Nd:YAG laser, the TALIF signal being acquired with a gated photomultiplier tube. In the following, a detailed description of each component of this setup is given (cf. Figure1).

2.2.1. <u>The laser system</u>

The ns-TALIF spectroscopy was performed using a tunable dye laser (SirahLasertechnik Cobra-Stretch) pumped by the second harmonic of a 500 mJ/pulse Nd:YAG laser (Spectra-Physics Quanta-Ray Lab-Series 170-10).The dye laser beam was subsequently processed by two BBO crystals, which yielded a laser beam with three harmonics that were separated using two fused silica Pellin-Broca prisms. The desired third harmonic that may be tuned between 204 nm and 207 nm was spatially selected by the Pellin-Broca prisms. A typical energy value obtained for this harmonic is 2.5mJ/pulse. The laser wavelength was systematically calibrated during LIF experiments using an iodine cell.

The obtained pure UV-C radiation was shifted vertically by means of a periscope and then focused with a 35 cm-focal-length lens at the center of the reactor chamber. In order to measure the laser energy, the beam exiting the chamber was collimated and collected with a calorimeter (Coherent J-10MB-LE), which was connected to an oscilloscope (Tektronix TDS 3052, 500 MHz, 5 GS/s).

2.2.2. <u>Fluorescence collection system</u>

One of the difficulties encountered with the investigated discharge is the very high level of the background emission obscuring the relatively weak TALIF signal.  The collection of N-atom fluorescence was therefore performed with a photomultiplier tube that allows for gated operations (Hamamatsu H11526-20-NF) and provides a linear regime up to 30 mA current. Furthermore, a focusing lens (f=10 cm) in a 4f configuration along with a 600 μm slit, and appropriate band-pass filters, were used to limit the probed volume, and the wavelength range to the fluorescence domain, respectively. With this configuration, the amount of background emission collected by the photomultiplier tube is suppressed, and the TALIF signal-to-noise ratio (S/N) is improved.

The anticipated time-resolved TALIF experiments require the synchronization of the microwave power pulse that ignites the plasma, the laser source that generates the fluorescence, and the gated photomultiplier tube that collects the fluorescence signal. For this purpose, the whole system was triggered using the visible component of the dye laser that was directed to a photodiode (PD1, ThorlabsSM05PD2A) connected to a 4 GHz-20 GS/s oscilloscope (Rohde & Schwartz RTO 1044). This oscilloscope converts the nanosecond signal received from the photodiode to a 1 ms signal that triggers a multifunction I/O device (National Instruments USB 6343, 500 kS/s, 1 MHz). This device generates the triggers for the plasma and the photomultiplier tube. Actually, for a single TALIF measurement, two laser pulses are needed: the first is to trigger the system and the second is actually needed to generate and monitor the fluorescence signal. Therefore, a delay is imposed between the triggering laser shot and the 5 μs photomultiplier-tube gate so as to capture the whole 400 ns fluorescence signal (c.f. figure 1). Similarly, the delay between the triggering laser shot and the microwave pulse is varied so as to explore the temporal evolution of the plasma.

The control of the whole platform, along with the signal acquisition and data processing was performed using a dedicated LabVIEW$^{TM}$ application.

A typical TALIF experiment performed in the frame of this work consisted in scanning the laser frequency ($v_l$) around the 206.67 nm and 204.13 nm for N-atom and Kr-atom, respectively, and acquiring for each value of $v_l$: (i) the time-resolved fluorescence signal $d^2s_f(v_l, t)$ recorded during a time-duration dt around t, i.e. during [t-dt/2, t+dt/2], generated by the laser radiation in the frequency range [$v_l$-d$v_l$/2, $v_l$+d$v_l$/2], (ii) the laser pulse energy signal $E_l$, and (iii) the iodine fluorescence signal used for wavelength calibration.

## 3. A simplified methodology for N-atom absolute density measurement

*3.1 N-atom density determination using krypton calibration*
We have used the TALIF scheme proposed by Bengsston et al. [10] that shows much larger sensitivity [11] than the very first scheme proposed by Bischel et al. [8]. In this scheme, N-atom in the $(2p^3)^4S^o_{3/2}$ ground state, denoted as N in the following, absorbs two photons produced by a laser pulse centred around 206.67 nm, and undergoes an electronic transition to the $(3p)^4S^o_{3/2}$ excited state hereafter denoted N$^*$. Assuming that the laser-atom interaction is dominated by the two-photon absorption, the density of the produced excited state is directly related to the offset frequency $\delta v_N = 2v_l - v_N$ where $v_l$ and $v_N$ are the laser central frequency and atomic transition frequency of the absorption line respectively. After the laser pulse, N$^*$ undergoes a first-order de-excitation through radiative decay and collisional quenching. The radiative de-excitation to the $(3s)^4P_{1/2, 3/2, 5/2}$ degenerate state produces three fluorescence lines at 742.364, 744.230 and 746.831 nm. Thus, during the de-excitation phase just after the laser pulse, N$^*$ density follows an exponential decay, with a time constant $\tau_{f,N} = \frac{1}{\sum_i A_{i,3p \to 3s} + Q}$, $A_{i,3p \to 3s}$ being the Einstein coefficient of the $i^{th}$ radiative de-excitation process [31] and Q being the rate of collisional quenching of N$^*$. The resulting incremental time-variation, $d^2n_{N^*}$, during dt, of the instantaneous density of the excited state, $dn_{N^*}(v_l \pm d(v_l), t)$, resulting from a laser pulse centred at a laser frequency $v_l \pm d(v_l)$, corresponding to an offset frequency $\delta v_N \pm d(\delta v_N)$, can be expressed as:

$$d^2 n_{N^*} = G^{(2)} \sigma_N^{(2)} n_N \left(\frac{I_l T(t)}{h v_l}\right)^2 g(\delta v_N) d(\delta v_N) dt - \frac{dn_{N^*}(v_l \pm d(v_l), t)}{\tau_{f,N}} dt \qquad (1)$$

Where $G^{(2)}$ and $\sigma_N^{(2)}$ are the two-photon statistical factor and the two-photon excitation cross-section ($m^4$) for the 2p→3p transition at a frequency $v_N$, respectively. $I_l$ and $T(t)$ are the radiant exposure (J. m$^{-2}$) and the normalized time profile (s$^{-1}$) of the laser pulse. $g(\delta v_N)$ is the statistical overlap factor between the nitrogen absorption line and the normalized laser line such that $\int_{-\infty}^{+\infty} g(\delta v_N) d(\delta v_N) = 1$. $g(\delta v_N) d(\delta v_N)$ also represents the laser energy fraction that is actually used to excite the atoms.

The measured fluorescence signal $d^2 s_{f,N}(v_l, t)_N$ during a time interval dt is proportional to the instantaneous density of the excited state $dn_{N^*}(v_l \pm d(v_l), t)$:

$$d^2 s_{f,N}(v_l, t) = K_g \eta_N \sum_i T_{i, 3p \to 3s} A_{i, 3p \to 3s} \, dn_{N^*} dt \qquad (2)$$

With $T_{i,3p \to 3s}$ the transmittance of the optics used for collecting the fluorescence signal and $\eta_N$ the sensitivity of the sensor over the spectral width of the fluorescence signal. $K_g$ is a geometrical factor that depends on several optical parameters that are difficult to estimate, e. g., the probed volume, the solid angle of the fluorescence volume subtended on the sensor. This factor shows the same value for N and Kr TALIF systems.

Substituting $dn_{N^*} dt$ by its value as a function of $d^2 s_{f,N}(v_l, t)$ as given by equation (2) in equation (1), and integrating over a long time period, one gets:

$$\frac{\int_{t=0}^{\infty} d^2 s_{f,N}(v_l, t)}{(E_l / h v_l)^2} = \frac{K_g}{S^2} \eta_N \sum_i T_{i, 3p \to 3s} A_{i, 3p \to 3s} \, G^{(2)} \sigma_N^{(2)} n_N \tau_{f,N} g(\delta v_N) d(\delta v_N) \int_0^{\infty} T^2(t) dt \qquad (3)$$

Where $E_l$ is the laser pulse energy recorded by the photodiode such that $I_l = E_l / S$, $S$ being the cross-section of the laser beam. The factor $d\Psi_N = \frac{\int_{t=0}^{\infty} d^2 s_{f,N}(v_l, t)}{(E_l / h v_l)^2}$ is the ratio of the number of detected fluorescence photons during dt to the square of the number of laser photons for a laser central frequency in the range $v_l \pm d(v_l)$.

In order to eliminate the unknown geometrical factors and to determine N-atom density, Krypton is used as calibrating gas as described by Niemi et al. [13]. A factor $d\Psi_{Kr}$ may also be defined for krypton and expressed as function of the Kr-atom density using an equation similar to (3):

$$d\Psi_{Kr} = \frac{K_g}{S^2} \eta_{Kr} T_{Kr} A_{Kr 5p \to 5s} G^{(2)} \sigma_{Kr}^{(2)} n_{Kr} \tau_{f,Kr} g(\delta v_{Kr}) d(\delta v_{Kr}) \int_0^{\infty} T^2(t) dt \qquad (4)$$

Where $T_{Kr}$ and $\eta_{Kr}$ are the transmittance and the sensitivity of the photomultiplier tube for the spectral range used for the detection of Kr-atom. $A_{Kr 5p \to 5s}$ is the radiative decay rate for the Kr 5p→5s transition observed [32].

Taking the ratio of equations (3) and (4) one may express the ground state N-atom density as a function of the known Kr-atom density:

$$n_N = \frac{\eta_{Kr}}{\eta_N} \frac{T_{Kr} A_{Kr 5p \to 5s}}{\sum_i T_{i,3p \to 3s} A_{i,3p \to 3s}} \frac{\sigma_{Kr}^{(2)}}{\sigma_N^{(2)}} \frac{\tau_{f,Kr}}{\tau_{f,N}} \frac{d\Psi_N}{d\Psi_{Kr}} \frac{g(\delta\nu_{Kr})}{g(\delta\nu_N)} n_{Kr} \qquad (5)$$

The ratio of the generalized two-photon excitation cross-sections $\sigma_{Kr}^{(2)}$ and $\sigma_N^{(2)}$, for the Kr $(4p^6)^1S_0 \to 5p'[3/2]_2$ and N $(2p^3)^4S_{3/2}^o \to (3p)^4S_{3/2}^o$, respectively, is taken from [13].

In equation (5), all the quantities except the ratio $\frac{g(\delta\nu_{Kr})}{g(\delta\nu_N)}$ are readily available after capturing the fluorescence generated by one laser pulse. The estimation of this ratio requires the knowledge of the profiles of the laser beam and the absorption lines of N-atom and Kr-atom. In particular, the absorption line profiles depend on the local plasma conditions and are not always easy to determine from the experiments.

This difficulty is generally circumvented by capturing the fluorescence signals generated by several laser pulses with different central frequency values $\nu_l$ around $0.5\nu_A (A = N \text{ or } Kr)$ and integrating $d\Psi_A$ over the whole laser central frequency range. As a matter of fact, if we consider the case of N-atom fluorescence, integrating equation (3) over the laser frequency domain yields the following relationship:

$$\int_{\delta\nu=-\infty}^{\infty} d\Psi_N = \frac{K_g}{S^2} \eta_N \sum_i T_{i,3p \to 3s} A_{i,3p \to 3s} \, G^{(2)} \sigma_N^{(2)} n_N \tau_{f,N} \int_0^{\infty} T^2(t) dt \qquad (6)$$

Similar equation can be derived for Kr-atom which is:

$$\int_{\delta\nu=-\infty}^{\infty} d\Psi_{Kr} = \frac{K_g}{S^2} \eta_{Kr} T_{Kr} A_{Kr 5p \to 5s} G^{(2)} \sigma_{Kr}^{(2)} n_{Kr} \tau_{f,Kr} \int_0^{\infty} T^2(t) dt \qquad (7)$$

Taking the ratios of equation (6) and (7), yields the following expression for the N-atom density:

$$n_N = \frac{\eta_{Kr}}{\eta_N} \frac{T_{Kr} A_{Kr 5p \to 5s}}{\sum_i T_{i,3p \to 3s} A_{i,3p \to 3s}} \frac{\sigma_{Kr}^{(2)}}{\sigma_N^{(2)}} \frac{\tau_{f,Kr}}{\tau_{f,N}} \frac{\int_{-\infty}^{\infty} d\Psi_N}{\int_{-\infty}^{\infty} d\Psi_{Kr}} n_{Kr} \qquad (8)$$

Equation (8), referred to as full excitation method (FEM), is widely used in the literature to determine N-atom densities by TALIF experiments. It implicitly captures the overlap factor without any assumption. However, its use requires the determination of the time and frequency integrated fluorescence signals for each density value measurements. In our experiments, this typically needs acquiring 70 spectral points which took approximately 20 minutes of acquisition. This is obviously not suitable for time-resolved measurements required in the context of the present work on pulsed plasma.

To circumvent this issue, we first make use of the specific discharge conditions investigated here in order to estimate $g(\delta\nu_A) d(\delta\nu_A)$, with A = N or Kr, and subsequently use equation (5) to readily determine N-atom density using a single laser shot at the central frequency of $0.5\nu_A$. The procedure used for this purpose is discussed in the following.

From the experimental point of view, $g(\delta v_A)d(\delta v_A)$ may be determined from the full excitation spectrum. As a matter of fact, taking the ratio of equations (3) and (6), one obtains for an atom A = N or Kr :

$$g(\delta v_A)d(\delta v_A) = \frac{d\Psi_A}{\int_{\delta v=-\infty}^{\infty} d\Psi_A} \quad (9)$$

Actually, $g(\delta v_A)d(\delta v_A)$ depends on the spectral profiles of the laser and the absorption lines. The spectral profile of the laser assumed in this study is Gaussian and this is also the case for the absorption lines in our conditions, since they are dominated by Doppler broadening for both N-atom and Kr-atom. In this case, the overlap factors may be easily expressed as:

$$g(\delta v_A)d(\delta v_A) = \frac{\sqrt{4ln2/\pi}}{\Delta v_{G,A}^{eff}} e^{-4ln2\left(\frac{\delta v_A}{\Delta v_{G,A}^{eff}}\right)^2} d(\delta v_A) \quad (10)$$

Where $\Delta v_{G,A}^{eff} = \sqrt{2\Delta v_{G,l}^2 + \Delta v_{G,A}^2}$ is the effective Gaussian Full-Width at Half Maximum (FWHM) due to convolution of the laser ($\Delta v_{G,l}$) and the transition profiles ($\Delta v_{G,A}$). $\Delta v_{G,A}^{eff}$ is experimentally obtained from the full excitation spectrum.

The laser broadening $\Delta v_{G,l}$ was determined from the TALIF signal of krypton at 300 K (without plasma). In such a condition, the Doppler broadening of the Krypton absorption line is estimated to be 0.13 cm$^{-1}$. The measured broadening of the krypton full excitation spectrum line is 1.59 cm$^{-1}$ which is much greater than the Doppler broadening of Krypton at 300 K. The resulting value of the laser line broadening is 1.10 cm$^{-1}$. It is worthy to mention that this value is significantly larger than those usually reported in the literature, for e.g. 0.25 and 0.36 cm$^{-1}$ in [14] and [21] respectively.

The determination of the absorption line broadening, $\Delta v_{G,N}$, for N-atom in the plasma requires the knowledge of the gas temperature. This was estimated from the rotational temperature of $N_2(C^3\Pi_u)$ electronically excited state, using optical emission spectroscopy (OES) applied to the transition $N_2(C^3\Pi_u \rightarrow B^3\Pi_g, \Delta v=-2)$ and for stationary operation of the MW plasma. The rotational temperature varied between 500 and 600 K for the investigated range of MW power. The resulting absorption linewidth for nitrogen remained almost constant at a value of approximately 0.46 cm$^{-1}$, which is two times smaller than the laser broadening.

In any case, since the gas temperature does not change significantly, the Doppler broadening of N-atom, and thus $\Delta v_{G,N}$ and $g(\delta v_N)$, should remain almost constant for the considered discharge parameters. This was confirmed by measuring the FWHM of the excitation spectrum $\Delta v_{G,N}^{eff}$ over a wide range of plasma conditions as shown in Figure 2.

This finding resulted in a significant simplification of the measurement procedure. As a matter of fact, the experiments to estimate the full integral $\int_{\delta v=-\infty}^{\infty} d\Psi$ have to be performed only once in order to obtain $g(\delta v_A)$. The N-atom density can then be readily determined through equation (5) using a single laser shot at a central frequency of $0.5v_A$, which significantly simplifies the measurement procedure. The use of the peak value at 0.5 $v_A$ enables achieving the maximum overlap between the laser and the absorption lines, thus limiting the uncertainties on the measured N-atom density. This method is referred to as Peak excitation method (PEM).

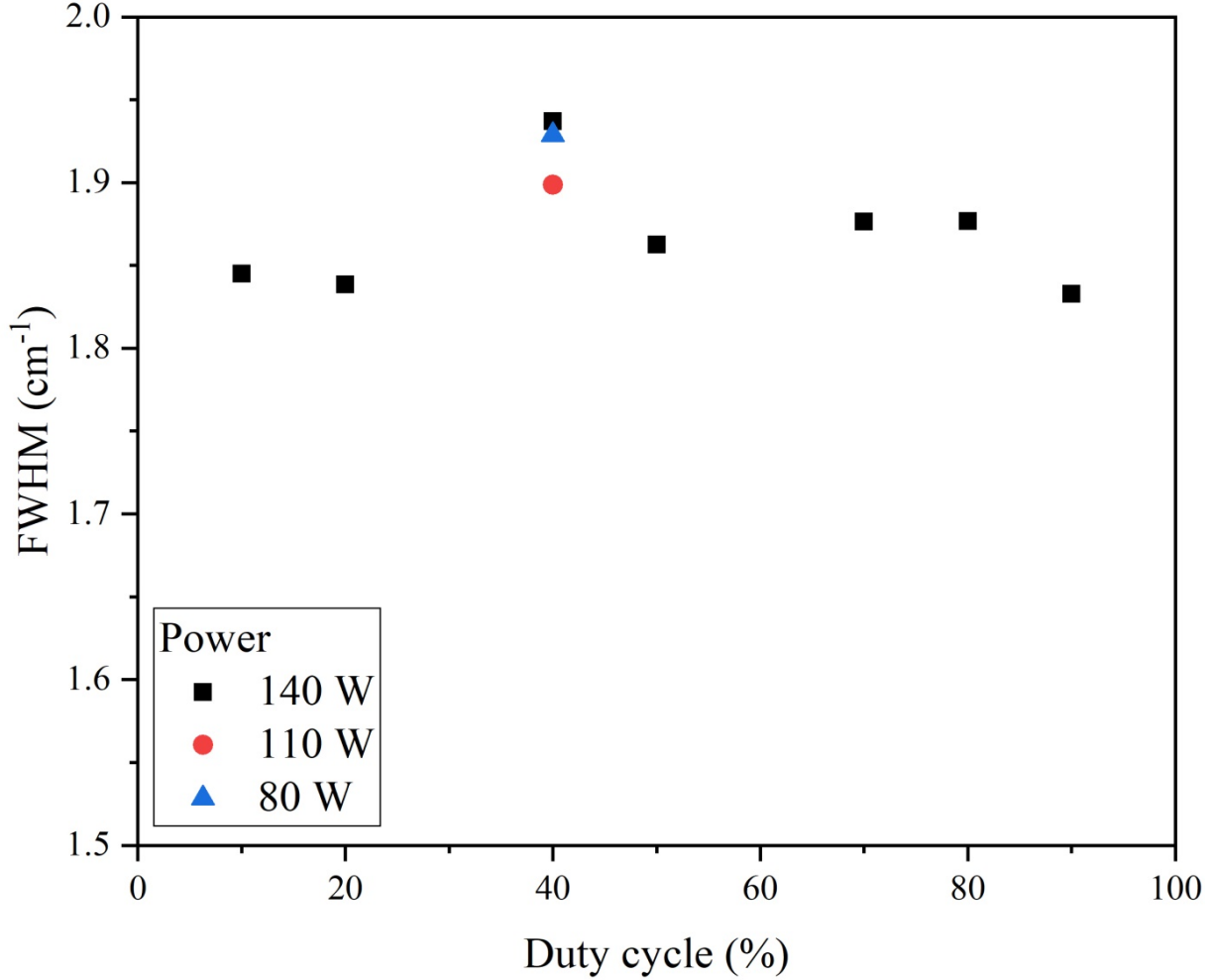

Figure 2: experimental spectral width in cm$^{-1}$, measured at the fundamental wavelength of the dye laser (16125.135 cm$^{-1}$), and deduced from the laser excitation spectra for several duty cycles and three MW powers.

*3.2 Signal Processing procedure*

The temporal integral $\int_{t=0}^{\infty} d^2 s_{f,N}(\nu_l, t)$ and $\tau_f$ are directly determined from the analysis of the time-resolved TALIF signal. The raw photomultiplier signal for nitrogen was composed of two parts (i) the time varying N* atom fluorescence and (ii) a plasma background emission identified as the N$_2$ FPS, i.e., N$_2$(B$^3\Pi_g \rightarrow$ A $^3\Sigma_u^+$, $\Delta v$ = -2), that was collected in the 728-750 nm wavelength range (this will be denoted 728-750 nm FPS emission). The overlap between these two components resulted in very poor signal-to-noise ratios, varying between 8 for the best case and 1.6 for most cases. The temporal integral of the raw signal would thus carry large uncertainties. To reduce these uncertainties, the photomultiplier signal was post-processed using a non-linear least square fit prior to performing the temporal integration. Figure 3 shows a typical raw photomultiplier signal along with its non-linear least-square fit. As indicated in figure 3, one may distinguish three different sections of the signal: (i) the constant plasma background region before the initiation of the laser pulse at a time $t_0$, (ii) the fast increase of the fluorescence signal during the laser pulse and (iii) the exponential decay with time constant $\tau_f$ after time $t_1$.

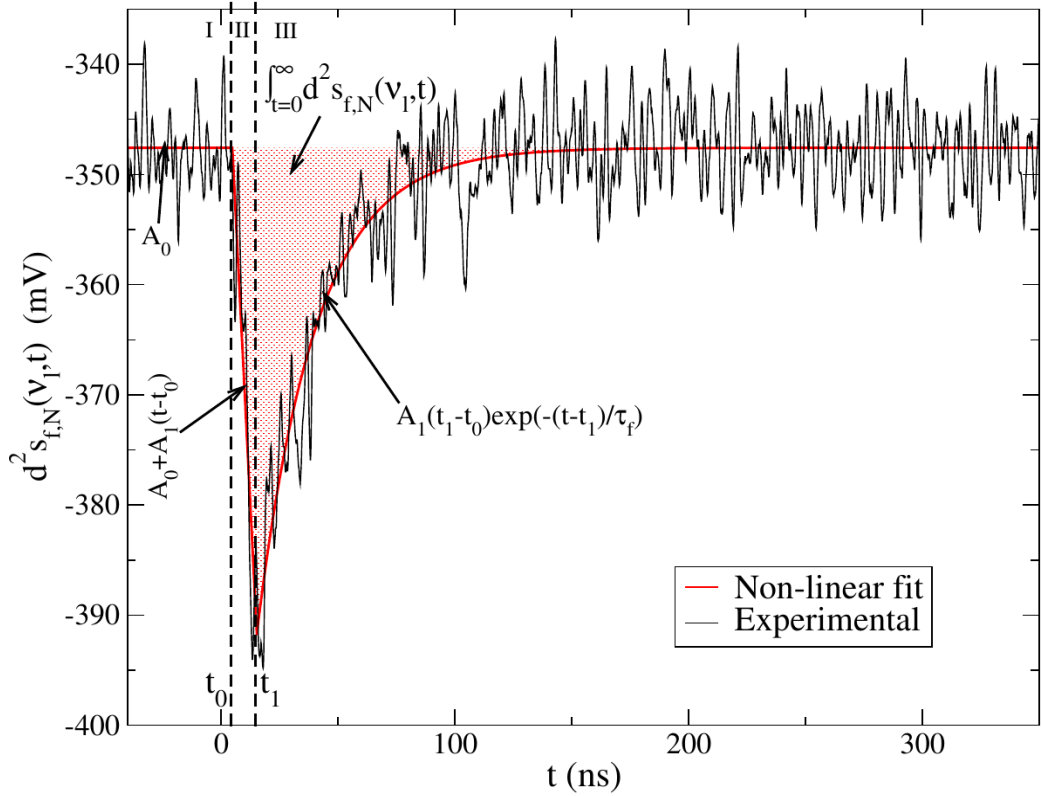

Figure 3: A sample raw photomultiplier signal (black line) and the corresponding fit, indicating the three regions, I, II and III, of the signal (as described in the text) and the corresponding temporal integral of the signal $\int_{t=0}^{\infty} d^2 s_{f,N}(\nu_l, t)$.

The fitting function is given by:

$$d^2 s_{f,N}(\nu_l, t) = \begin{cases} t < t_0: & A_0 \\ t_0 < t < t_1: & A_0 + A_1(t - t_0) \\ t > t_1: & A_1(t_1 - t_0) exp\left(-\frac{t - t_1}{\tau_f}\right) \end{cases} \quad (11)$$

All the parameters of the fit ($\tau_f, A_0, A_1, t_1$ and $t_0$) are obtained simultaneously. Thus, the time-integrated signal (filtered) for N-atom can be written as follows :

$$\int_{t=0}^{\infty} d^2 s_{f,N}(\nu_l, t) = -\tau_f A_1(t_1 - t_0) - \frac{1}{2} A_1(t_1 - t_0)^2 \quad (12)$$

$\int_{\delta\nu_N=-\infty}^{\infty} d\Psi_N$ was determined from the laser excitation spectral profile of $d\Psi_N(\delta\nu_N)$. For this purpose, the fitting procedure was applied to a set of 70 raw photomultiplier signals (cf. figure 3) resulting from the fluorescence generated by 70 laser pulses with central frequencies varying between 96743 cm$^{-1}$ and 96754 cm$^{-1}$ (103.35-103.37 nm). The as determined laser excitation spectral profiles of $d\Psi_N(\delta\nu_N)$ is shown in Figure 4. For the experimental conditions of the study, the normalized spectral line profile $d\Psi_N / \int_{\delta\nu_N=-\infty}^{\infty} d\Psi_N$, which is also $g(\delta\nu_N)d(\delta\nu_N)$, can be well approximated by a Gaussian profile (also shown in Figure 4) as expressed in equation (10).

The efficiency of the least-square fit for determining the time-integral can be clearly seen by comparing the very noisy raw spectral profile, also shown in Figure 4, with the spectral profile obtained using the fitting procedure given by equation 11.

A similar procedure was applied to determine the LIF laser excitation spectrum of $I_2$ used for laser wavelength calibration. This is also depicted in Figure 4 that shows that the spectrum broadening extends over several $I_2$ fluorescence maxima, which indicates a fairly high broadening of the laser line, typically 2 cm$^{-1}$.

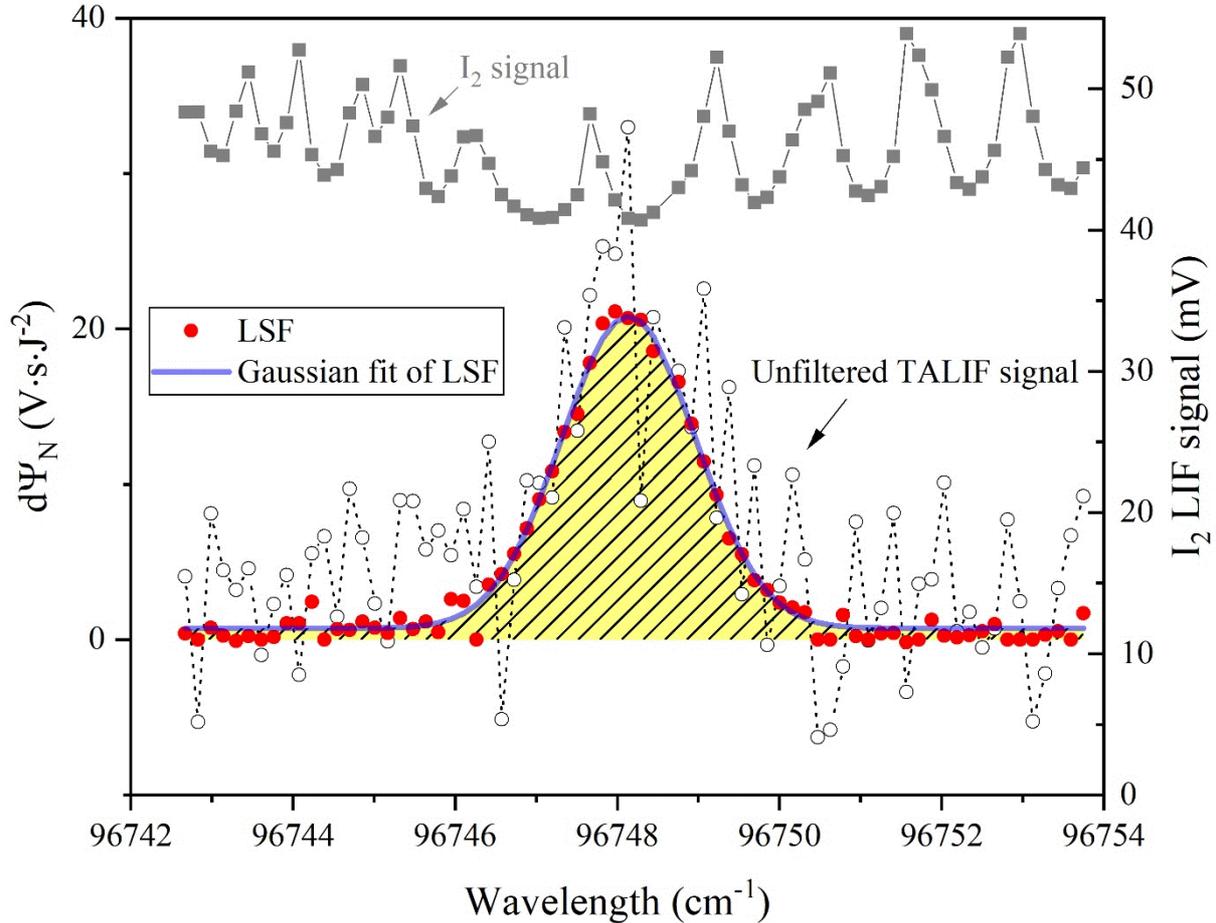

Figure 4: N-atom TALIF laser excitation spectra obtained from raw data (black opencircles), and data processed using the nonlinear least-square fitting procedure (red solid circles) and their Gaussian fit (blue line). The pressure is equal to 20 Pa and the maximum MW power is equal to 140 W. The $I_2$ LIF laser excitation spectrum (grey solid squares) used for absolute calibration of the laser wavelength is also plotted above.

*3.3 Fluorescence decay time measurements for Kr and N atoms*

The determination of N-atom density (equations (8) or (10)) requires determining the ratio of the fluorescence decay times of krypton and nitrogen. The decay time for the krypton fluorescence line intensity at 587.09 nm was measured at 10 Pa and a value of 31.0 ± 1.2 ns was found. In fact under the low pressure conditions considered, the quenching is negligible and the fluorescence decay time is almost equal to the 5p'[3/2]$_2$ radiative lifetime. Using the quenching rate-constant values given in [13] we estimated a value of 31.8 ± 1.3 ns for this radiative lifetime. This value is slightly lower than those previously reported in [13], i.e. 34.1 ns, and [33], i.e. 35.4 ns, but larger than the value of 26.9 ns given in [34].

The fluorescence decay-time for N($^4$S) atoms was measured at 20 Pa and was found equal to 26.0 ± 1.4 ns for all the investigated plasma conditions (cf. Figure 5). Using the rate-constant values given in [11] and [13] for the quenching of N($^4$S) by $N_2$-molecules, we found a value of 26.2 ± 2 ns for the N(3p⁴S⁰) radiative life time in agreement with [10,11,15]. As expected, the collisional quenching by $N_2$-molecules is negligible and does almost not reduce the fluorescence decay-time.

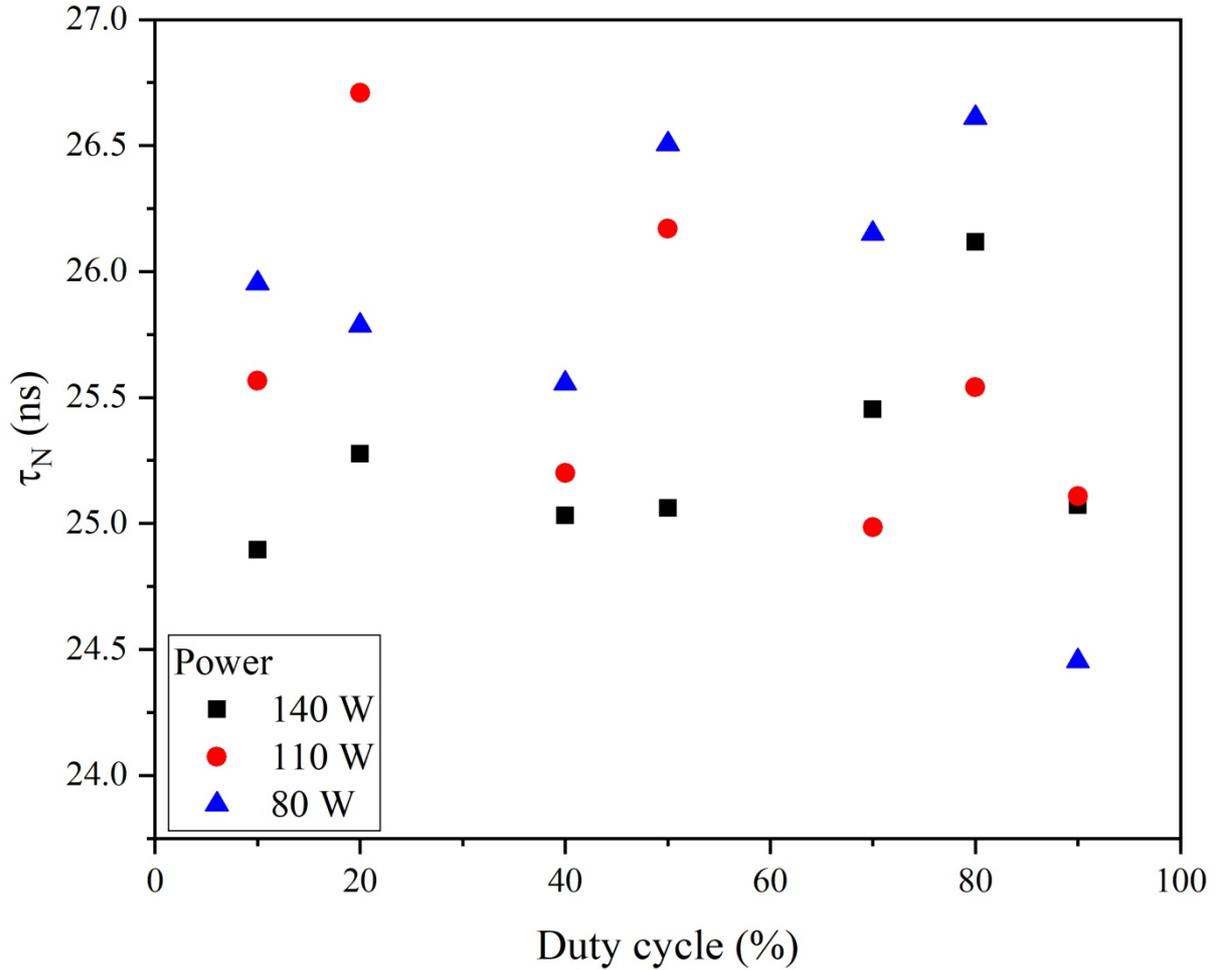

Figure 5: Decay time of N-atom fluorescence as a function of the duty cycle for different average input power values. The mean value of N($^4$S) fluorescence decay time estimated from all these data was assessed to be 26 ± 1.4 ns.

*3.4 Uncertainty analysis*

The uncertainty on the measured nitrogen atom densities was calculated using the propagation of uncertainties of the independent variables [35]. The uncertainties of the TALIF integrals (d$\Psi$ and $\int_{\delta v=-\infty}^{\infty} d\Psi$), $\Delta v_{G,A}^{eff}$ and $\tau_f$ were obtained directly from the non-linear fitting procedures [36]. The other independent variables were either obtained from the literature (two photon absorption cross-sections) or datasheets of the manufacturers (transmission of filters, photomultiplier sensitivity). Figure 6 shows the contributions of the different independent variables to the uncertainty of the nitrogen atom density. The largest contribution to the relative error on nitrogen atom density comes by far from the uncertainty on the ratio of the two-photon absorption cross-sections for nitrogen and krypton, 50%, [13]. The second error source comes from the uncertainty on the krypton radiative lifetime, 20% [32]. The uncertainties on the time-integrated peak fluorescence intensity, d$\Psi$, determined using the procedure described above contributes to 10-15% to the global relative error on the density. The contributions of the other error sources are below 5%.

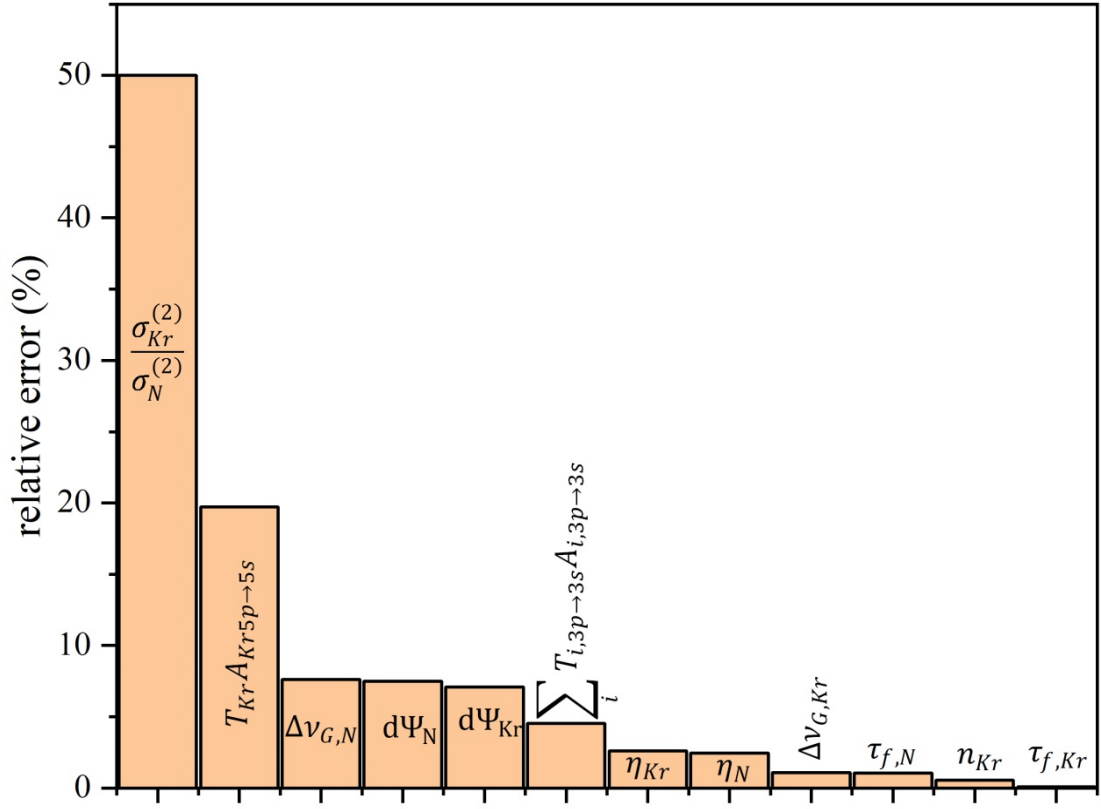

Figure 6: relative errors for the different terms of the equation (5)

*3.5 Power dependence*
The different standard procedures used to determine atomic densities from TALIF measurements are generally valid only if the laser instantaneous and local power density remains low enough to avoid additional laser-atom interaction phenomena such as field coupling, photoionization, photodissociation and stimulated emission [37,38]. The analysis of the TALIF signal intensity dependence to the laser energy, i.e., the so called power dependence or PD factor [39], makes it possible to check if the experimental conditions insure a laser-atom interaction dominated by the processes involved in the TALIF scheme. A value of 2 for this factor, i.e., a quadratic regime of excitation, indicates that the fluorescence dominates the laser-atom interaction processes.

From the experimental point of view several methods may be used to vary the laser energy in order to determine the PD value. Unfortunately, there is no clear conclusion on the best one. Therefore, in this work, we tested three approaches. In the first one the energy was changed using a half-wave plate and a Glan-Taylor polarizer. In the second one, the energy was varied by modifying the angle of the second BBO crystal in the dye laser. The third one was based on the use of a set of neutral density (ND) filters.

The method based on Glan Taylor polarizer is very sensitive to the beam polarization, which resulted in repeatability issue since it was very difficult to inject the laser beam at the Brewster angle in the Pellin-Broca prism. As for the second method, the laser shape changed significantly with the energy, which prevented using equation (1) that assumes a constant beam shape. The last method was in fact the most reliable and straight forward. It yielded a power dependence of ~2 for both nitrogen TALIF at 20 Pa and krypton TALIF at 1 Pa over the laser energy range 100-300 µJ (cf. Figure 7).

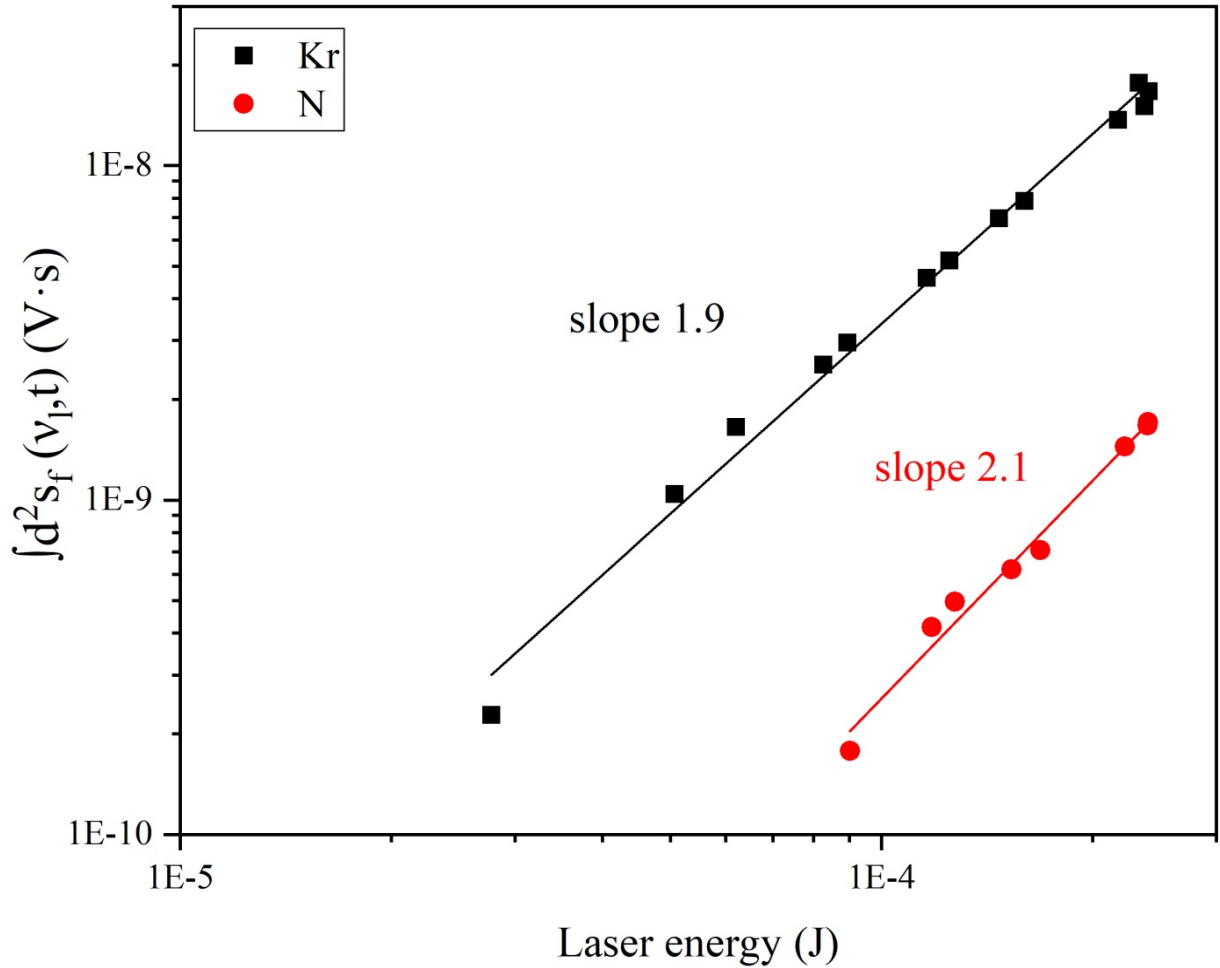

Figure 7: Measured variations of the time-integrated TALIF signal intensities as a function of the laser energy for nitrogen and krypton. For these measurements, the laser frequency was set at the peak of absorption ($\nu_l = \nu_A/2$).

## 4. Results and discussion

The plasma source was pulsed at the laser frequency of 10 Hz and the pressure was fixed at 20 Pa. The pulsed plasma regime is characterized by high- and low-power phases. Actually a zero input power during the low-power phase prevents the plasma re-ignition at the start of the high-power phase. Thus, it was mandatory to use an absorbed MW power value of 20 W during the low-power phase. In comparison, the absorbed power during the high-power phase was set to 80, 110 or 140 W. Under such power conditions, the volume of the highly emissive plasma zone, where a major part of MW power is likely to be dissipated, did not change significantly. In this study, we made measurements at a single position, 10 mm below the magnet (cf. Figure 1). Actually, the N-atom diffusion time over a characteristic length of 10 mm under a pressure of 20 Pa is below 1 ms, which is much smaller than the plasma pulse duration. Thus, this would result in approximately spatially homogeneous N-atom density that may be characterized by a single spatial measurement.

*4.1. Comparison between PEM and FEM approaches*

Figure 8 shows the variation of N-atom densities measured using the full excitation method (FEM) and the peak excitation method (PEM) as a function of the power coupled to the plasma. The considered discharge conditions correspond to a duty cycle of 0.4. The measurements were performed 1ms after the plasma was switched-off. The results obtained by the two approaches are in a very good agreement in the range 6-7 x $10^{18}$ m$^{-3}$. Hence, we subsequently employed the straightforward PEM to determine the N-atom density.

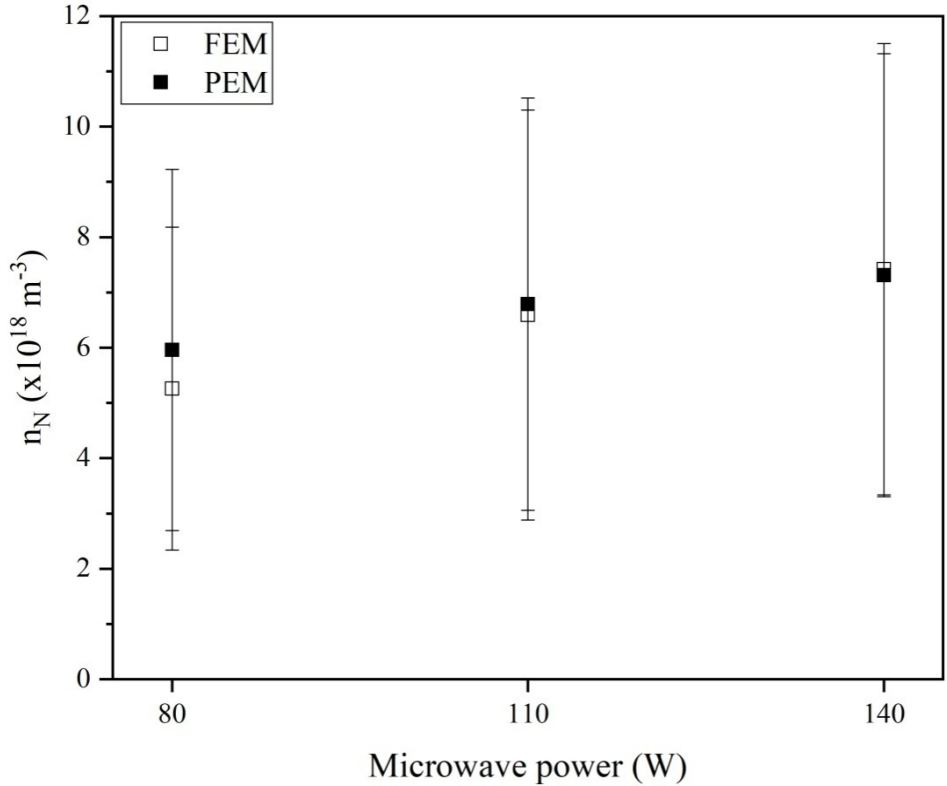

Figure 8: comparison of the atomic nitrogen density obtained using the full excitation method (FEM) and the peak excitation method (PEM) for three MW power values of the high power phase. The measurements were performed 1 ms after the plasma was switched-off. The pressure is equal to 20 Pa and the duty cycle is equal to 0.4.

*4.2. N-atom density measurements during the high power phase*

Figure 9 shows the variations of the stationary values reached by N atom-density and the 728-750 nm FPS emission intensity during the high power phase as function of the MW power deposited during this phase for a pulse period and a duty cycle of 100 ms and 0.5, respectively. Both are seen to increase similarly and linearly with the MW power.

In fact, for the considered value of duty cycle and pulsed period, the plasma reaches a stationary state during the pulse (see section 2. 3 and figure 10). The electron temperature is therefore mainly determined by the balance between the electron production and loss terms and does not depend on the deposited power [40,41]. The increase in the 728-750 nm FPS emission and the N-atom density with the power can only be explained by the increase in the electron density which is proportional to the power density if the plasma composition does not change significantly [40,41]. Thus, for the current experiments, we have a constant electron temperature, and the N-atom density and the FPS emission intensity would be proportional to the electron density and therefore to the power density. Since the measurements reported in figure 9 show that the N-atom density and the FPS emission intensity are proportional to the power, we can conclude that the power density is proportional to the absorbed power, which means that the plasma volume remains almost constant. We can also conclude that, despite the very emissive nature of the plasma, the methodology developed in this work enables predicting consistent trends of N-atom densities with MW power and FPS emission.

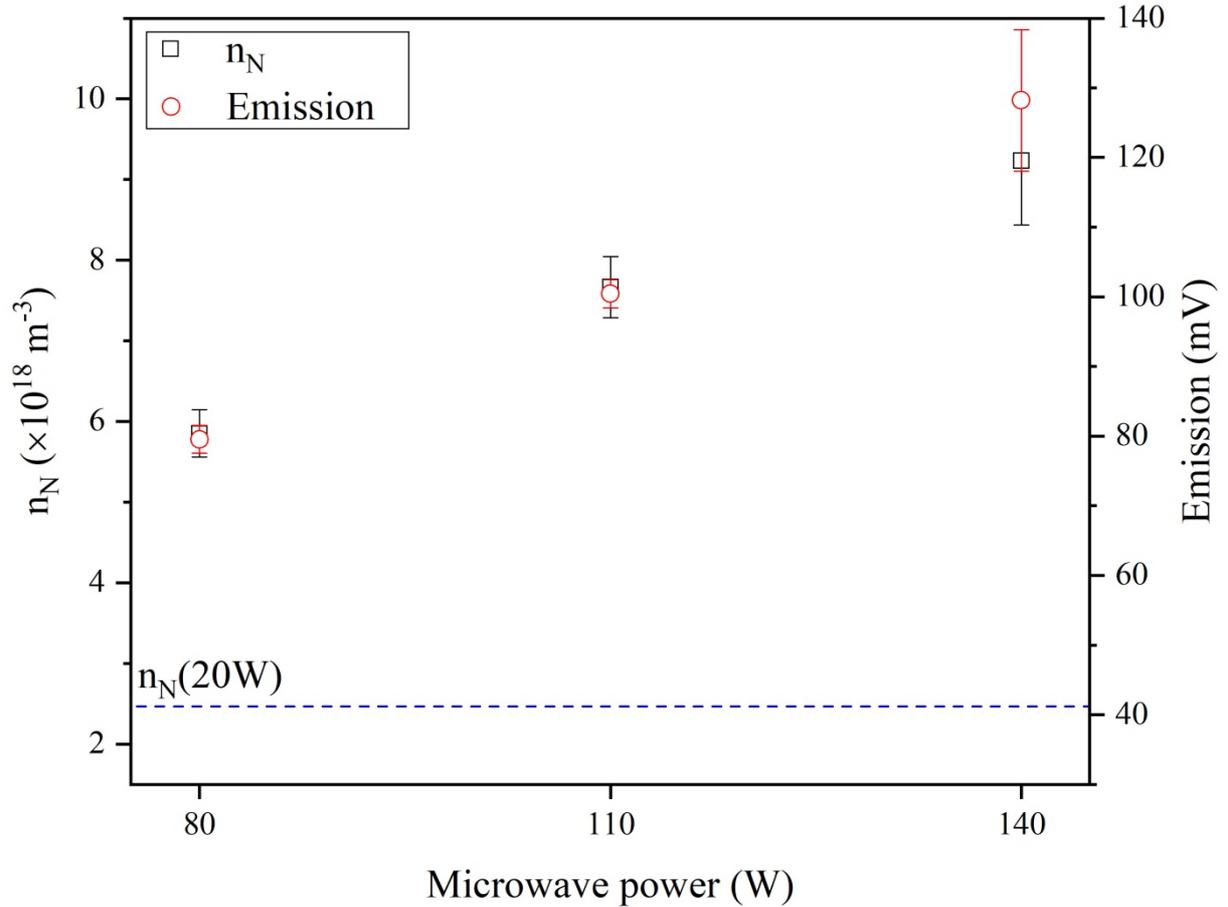

Figure 9: N-atom densities deduced from TALIF measurements (black open squares) and 728-750 nm FPS emission intensity (red open circles) during the high power phase as a function of the coupled MW power.

*4.3 $N(^4S)$ time-resolved density measurement and kinetic analysis during the transients*
The time-evolution of the N-atom density, the 728-750 nm FPS emission intensity, and the MW power are shown in figure 10. In this figure, the time origin, t = 0 s, is chosen at the start of the power increase (a) or decrease (b) at the transition between the high- and low-power phases. The variation of the 728-750 FPS emission intensity is indicative of the change in the population of the $N_2(B^3\Pi_g)$ electronic state (denoted as $N_2(B)$ in the following).
Also shown in the figure are the time-variation fits using first order exponential growth and decay functions that are used to estimate the rise and decay characteristic times. The rise and decay times of the absorbed MW power are approximately 0.45 ms and 0.07 ms, respectively. These are much smaller than the rise and decay times of the $N_2(B)$ emission intensity and the $N(^4S)$-atom density. Also, the measured plasma characteristics reach steady-state values during both high- and low-power phases. This indicates that the characteristic times for the governing processes of $N(^4S)$-atom and $N_2(B)$ kinetics are much smaller than the 100 ms pulse period.

The dynamics of the electron heating in pulsed plasmas is very fast and the electron temperature usually closely follows the time-scale of the absorbed MW power [42] (see also the discussion and figures at the end of this section). The sudden increase in the MW power should result in a large electron heating that leads to enhanced excitation, ionization and dissociation kinetics. Figure 10(a) shows that $N(^4S)$-atom density and $N_2(B)$ emission exhibit similar characteristic rise times, i.e., around 1 ms, which would therefore represent the dissociation and excitation characteristic times. Since the gas temperatures is fairly low, i.e., $T_g$=500-600 K, the excitation is governed by electron impact processes (R1) while the dissociation involve both an electron impact process on the ground state (R2) and the well-known vibrational excitation/relaxation route (R3) which also requires vibrational excitation by electron impact. Thus the time variation of the N($^4$S)-atom density and the $N_2$(B) emission intensity are also indicative of electron density build-up and ionization time-scale.

$$e + N_2(X^1\Sigma_g^+) \rightarrow e + N_2(B^3\Pi_g, v) \tag{R1}$$
$$e + N_2(X^1\Sigma_g^+) \rightarrow e + N(^4S) + N(^4S \text{ or } ^2D) \tag{R2}$$
$$e + N_2(v) \rightarrow N_2(w>v) + N \text{ or } N_2 \rightarrow N(^4S) + N(^4S) + N \text{ or } N_2 \tag{R3}$$

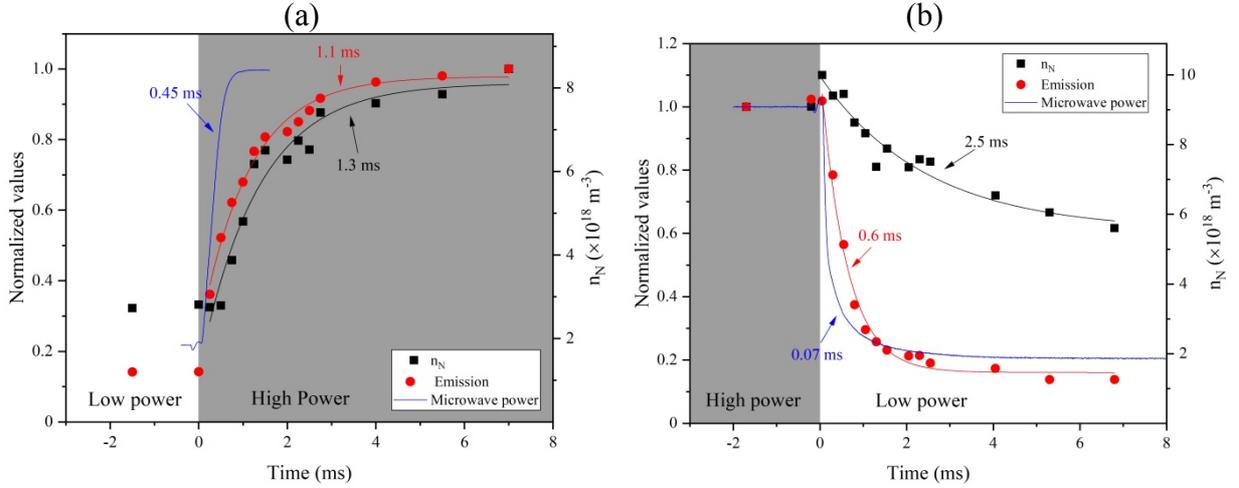

Figure 10: Temporal evolution of the N($^4$S)-atom density, the $N_2$(B) emission intensity and the MW power for (a) the high power phase and (b) the low power phase at pressure of 20 pa and 50% duty cycle. The time origin, t = 0, i.e. when transitioning between the high- and low-power phases, is chosen at the start of the power increase (a) or decrease(b). Time-resolution of these measurements is 250 µs.

As for the decay phase, since the dynamics of the electron heating is very fast, one would expect the electron temperature drop to closely follow the MW power variation. Therefore the high energy threshold electron impact excitation, ionization and dissociation processes are very likely to be rapidly quenched during the power decay phase (see also the discussion at the end of this section).
If the loss process of $N_2$(B) was mainly through radiative de-excitation, which has a characteristic lifetime of 10 µs [43], the decay of the $N_2$(B) emission would have closely followed the MW power decay. This is obviously not the case in this experiment which showed a much longer decay time of 600 µs for the FPS emission. Actually, $N_2$(B) kinetics probably follows a complex mechanism similar to that observed in a short-lived afterglow [44–46]. In that case, the kinetics of $N_2$(A), $N_2$(B) and $N_2$(C) are strongly coupled in a collisional-radiative loop involving several processes that are reported in [47]. The major exit channels from this loop are the processes:

$$N_2(A) + \text{wall} \rightarrow N_2 \tag{R4}$$
$$N_2(A) + N_2(a') \rightarrow N_4^+ + e \quad k=10^{-11} \text{ cm}^3.\text{s}^{-1} \text{ [48]} \tag{R5}$$
$$N_2(A) + N_2(A) \rightarrow N_2(X,v=0) + N_2(B \text{ or } C) \quad k=7.7 \times 10^{-11} \text{ cm}^3.\text{s}^{-1} \text{ [48]} \tag{R6}$$
$$N_2(A) + N(^4S) \rightarrow N_2(X,v=6-9) + N(^2P) \quad k_4=4 \times 10^{-11} \text{ cm}^3\text{s}^{-1} \text{ [47]} \tag{R7}$$

$$N_2(B) + N_2(X) \rightarrow N_2(X,v=0) + N_2(X) \quad k_3=0.15 \times 10^{-11} \text{ cm}^3\text{s}^{-1} \text{ [47]} \quad \text{(R8)}$$

If we assume a large de-excitation probability of $N_2(A)$ at the wall, the diffusion/wall-de-excitation of this metastable should be diffusion limited. In this case, the $N_2(A)$ de-excitation characteristic time, i.e., reaction (R4), would correspond to the diffusion characteristic time of $N_2(A)$ that is given by $\Lambda^2/D$ where $\Lambda$ is the plasma characteristic length and D the diffusion coefficient. Assuming the same value for the diffusion coefficients of all $N_2$ electronic states and a Lenard-Jones $N_2$-$N_2$ interaction potential, a value of 0.03 m$^2$/s was estimated for D at a pressure of 20 Pa and a temperature of 600 K. This yields a 5 ms diffusion characteristic time for the 2 cm length scale estimated from the volume-to-surface ratio in our system. This diffusion time-scale is much longer than the observed decay time and we can conclude that reaction (R4) does not dictate the loss mechanism of electronic excited states of $N_2$. The estimation of the characteristic times of reactions (R5) and (R6) is more difficult since it requires the knowledge of the $N_2(A)$ and $N_2(a')$ densities. A lower limits for these time scales may be however calculated by noticing that the relative densities of these excited states with respect to $N_2$ ground states are always below 10$^{-3}$ [47,48]. Using this value of relative density and the rate constants given in [48] we can estimate lower limits of 42 ms and 5 ms for the characteristic times of reaction (R5) and (R6), respectively. These values are also much larger than the observed sub-millisecond decrease of the FPS emission intensity. These processes do not result in a depletion of the excited triplet state that is consistent with the one observed for the FPS emission during the early stage of the low power phase. Using the N-atom density measured in the present study, i.e., $n_N \approx 10^{19}$ m$^{-3}$, a characteristic time of 2.5 ms was estimated for reaction (R7) that is also not fast enough to explain the time decrease of the FPS just after the transition to the low power phase. Eventually, a value of 275 µs was estimated for reaction (R8). It appears therefore that in our condition, reaction (R8) is the major loss mechanism of excited electronic states in our discharge conditions. The characteristic time inferred from this reaction rate constant is not very far from the one measured experimentally, i.e., 600 µs. Taking into account the qualitative nature of our reaction timescale analysis as well as the uncertainty of the rate constants value of (R8) [47] one can even consider that the agreement is satisfactory.

The characteristic time of N-atom density decay is longer than that of the plasma emission. It was estimated to be 2.5 ms. Over such a long-time scale, it is usually assumed that the consumption of N-atom takes place through surface recombination processes at the reactor walls:

$$N(^4S) + \text{wall} \rightarrow 0.5 N_2 \quad \text{(R9)}$$

Surface recombination is generally a first order reaction which is controlled by the balance between the diffusion and surface recombination at the wall:

$$\frac{D}{\Lambda}(x_g - x_w) = -\frac{\gamma}{4} v_s^* x_w \quad (11)$$

Where D is the diffusion coefficient, $\Lambda$ is the plasma characteristic length, $\gamma$ is the sticking coefficient and $v_s^*$ is the mean thermal velocity. Further, $x_g$ and $x_w$ are the N-atom mole-fractions in the gas phase and at the reactor wall, respectively. Assuming a Lennard-Jones N-$N_2$ interaction potential, a value of 0.15 m$^2$.s$^{-1}$ was estimated for the diffusion coefficient of N-atom at 20 Pa and 600 K. This yields a value of 3.4 ms for the diffusion characteristic time, which is very close to the decay time seen in our experiments. This shows that the measured depletion kinetics of $N(^4S)$ atoms agrees with diffusion limited recombination process. In fact if we assume that the recombination process is limited by the surface reaction, the value of $\gamma$ determined from the measured decay time would be approximately $10^{-2}$ which is significantly larger than the value of $10^{-3}$ reported by [27] for stainless steel. This leads to the conclusion that the recombination kinetics is very likely to be limited by the diffusion process. Therefore the agreement between the measured decay time and estimated diffusion time would tend to validate our $N(^4S)$-density measurements. Actually, the $N(^4S)$ loss kinetics involves much more complex kinetics than a simple diffusion/recombination process. As a matter of fact, looking at figure 10b, one can notice a surge in the N-atom density just at the transition between the high and low power phases despite the poor time resolution in this figure. This reproducible surge is similar to the one observed in [17] where the authors showed an unexpected increase of $N(^4S)$ TALIF signal at the start of the plasma off phase in a pulsed microwave plasmas. The methodology developed in the present work provides TALIF measurements with a time resolution as low as 5 μs and a straightforward signal processing which enables investigating short time-scale kinetics effects for $N(^4S)$. This has been used to confirm and further study the existence of this fast increase in N-atom TALIF signal that is expected to occur within 500 μs during the early stage of the low power phase. Figure 11 shows time-resolved measurements with a much better resolution, i.e., 10 μs resolution, of the $N(^4S)$-atom density and the background plasma emission near the transition between the high- and low-power phases. This figure confirms the existence of a significant increase in the N-atom density (more than $10^{18}$ m$^{-3}$, which represents more than 10% relative variation) just after the transition to the low power phase. Actually, a high $N(^4S)$ density level is maintained during almost 1 ms after the power transition. Note that such behaviour is not observed for the plasma emission intensity that decreases sharply just after the transition.

The increase in the $N(^4S)$ density can be only due to the change in the kinetics of very short characteristic-time processes. This may involve either a decrease in the rates of N-atom consumption processes or an increase in the rates of N-atom production processes at the transition between the high- and low-power phases. Therefore, beside reactions (R2), (R7) and (R9), we considered the following gas phase processes that are reported in the literature [47–49] as far as $N(^4S)$ kinetics is concerned :

$$N(^2D) + N_2 \rightarrow N(^4S) + N_2 \tag{R10}$$
$$N(^2D \text{ or } ^2P) + \text{wall} \rightarrow N(^4S) \tag{R11}$$
$$e + N_2^+ \rightarrow N + N \tag{R12}$$
$$e + N(^2P \text{ or } ^2D) \rightarrow e + N(^4S) \tag{R13}$$
$$e + N(^4S) \rightarrow e + N(^2P \text{ or } ^2D) \tag{R14}$$

As mentioned previously (see also the discussion at the end of this section), the rapid transition from the high - to the low-power phases will first result in a decrease of the electron temperature, which immediately affects the rates of the electron-impact processes, and more particularly the ionization and dissociation processes. Then, the time-evolution of $N(^4S)$ is governed by the interplay of all the process mentioned previously, i.e., (R2)-(R3), R(7) and R(9)-(R14).

To evaluate the relative pre-dominance of these channels in the evolution of the $N(^4S)$-atom density, we made use of a self-consistent $N_2$ plasma model that solves for the two-term expansion electron Boltzmann equation coupled to the balance equations of vibrational states of $N_2$ ground state (46 levels), $N_2$ and N electronically excited states that results from collisional-radiative models of $N_2$ (18 states) and N (26 states), as well as the total energy equation that yields the gas temperature. The detailed description of this model, that makes use of the methodology described in [42] and collisional radiative models similar to those described in [47,49], is out of the scope of this paper. Here, we limit ourselves to the use of this model in order to analyse the time-variation of the electron density, the electron temperature, the $N_2$ vibrational distribution and the N-atom densities in order to evaluate the relative predominance of the different processes involved in $N(^4S)$ kinetics.

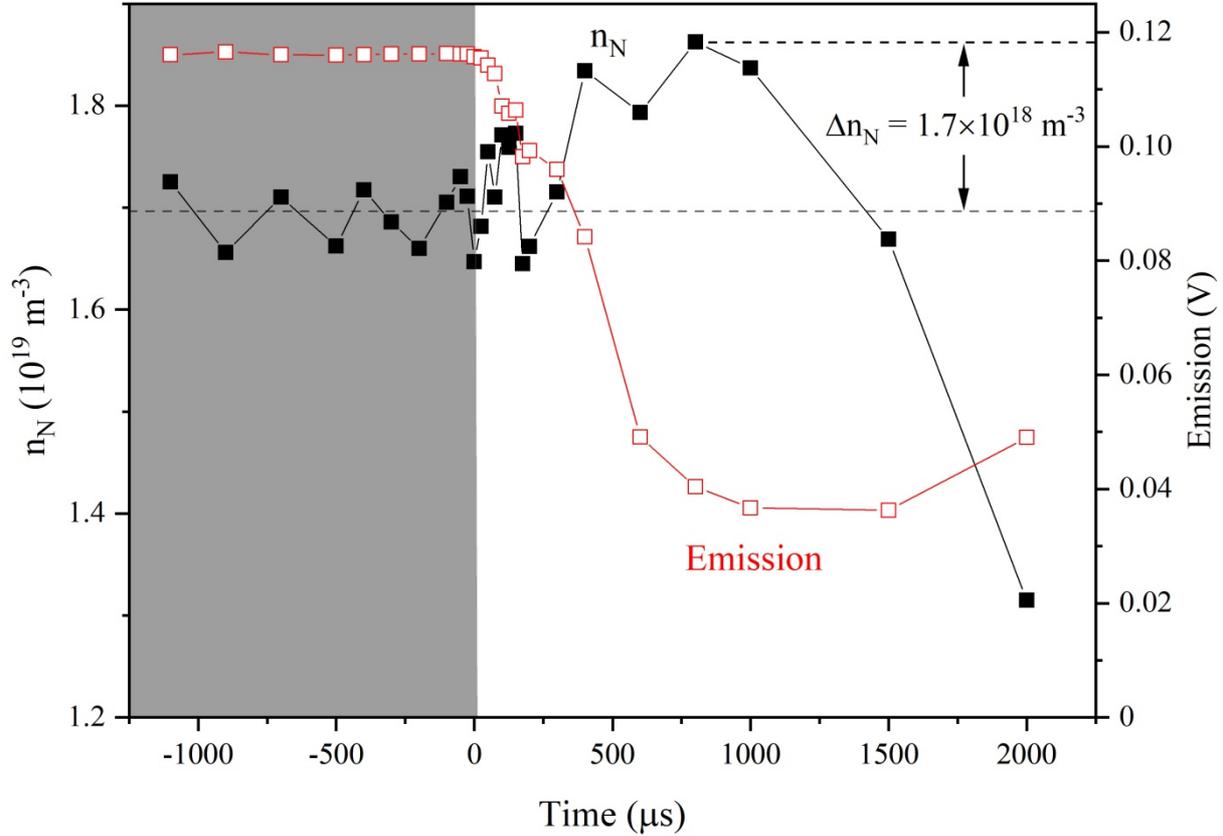

Fig 11: Time-resolved measurement of N-atom (black circles, left axis) and $N_2(B^3\Pi_g)$ emission (red squares, right axis) around the transition between the high and low power phases and during the early stage of the low power phase.

Figure 12 shows the time-variation of the calculated electron-temperature ($T_e$) and electron density ($n_e$) at the transition between the high- and low-power phases. We can see and confirm the previously mentioned and expected very fast and strong electron cooling phase during which the electron temperature decreases from 17000 K to 7000 K. This first decrease in $T_e$ is due to the fact that the very large inelastic collision frequency results in electron-heavy species energy transfer kinetics that are much faster than electron recombination kinetics. As a result, when the power is decreased, a similar electron population has to share a much smaller amount of energy. This results in the observed decrease in the electron temperature. Due to the very low $T_e$-value at the end of the very short cooling phase, the electrons start recombining and their population decreases gradually as may be seen in figure 12. This results in the observed subsequent increase in the electron temperature that reaches a steady state low power phase value of 15000 K, which is just 2000 K below the high power phase value. It appears therefore that the calculated $n_e$ and $T_e$ transient-duration of 250 μs are coupled. They are governed by the electron kinetics and more precisely by the balance between the ionization and electron loss processes. These later are dominated by wall recombination.

Figure 13 shows the calculated time-variation of the vibrational distribution functions (VDF) at different stages of the transient between high- and low-power phases. It is worthy to mention that the y-axis scale is extended for the low population value in order to emphasize the relevant population changes during the transition. The relative density of high-lying vibrational states (v>34) strongly decreases during the early low-power phase. A detailed rate analysis shows that this decrease benefits to the low-lying vibrational states through vibrational-de-excitation process, the contribution towards vibrational dissociation during the 500 µs being negligible (see Figure 14). Note that the total variation of relative density of the high vibrational levels is below $10^{-5}$ which is much below the observed N($^4$S) relative density increase that is around $10^{-3}$.

The time-variation of the rates of the processes involved in the kinetics of N($^4$S) are shown in figure 14 confirming that the contribution of process (R3), i.e., v-d mechanism, to N($^4$S) kinetics is very weak. Actually, during the high power phase, the kinetics of N($^4$S) is essentially governed by the balance between (i) the production through electron-impact dissociation of N$_2$ (reaction (R2)) and the de-excitation of N($^2$D/$^2$P) at the wall (reaction (R11)) and, to a lesser extent, by super-elastic collisions (reaction (R13)), and (ii) the loss through electron-impact excitation of N($^4$S) to N($^2$D) and N($^2$P) states (reaction (R14)) and wall-recombination (reaction (R9)). When the power is decreased, the electron cooling affects this balance through a strong decrease in the electron-impact dissociation of N$_2$, the N($^4$S)-atom excitation and, to a lesser extent, the N($^2$D/$^2$P) de-excitation. As a result, the kinetics of N($^4$S)-atom becomes dominated by the wall de-excitation of N($^2$D/$^2$P), as a source, and wall recombination, as a loss term, during the transient. This results in a net production of N($^4$S) during 200 µs, which explains the observed short time scale increase in N($^4$S) density in both the experiment and simulation (not shown). Actually, the surface recombination rate increases by approximately 10% during the early low power phase (which can be easily seen in the insert in figure 14). This indicates a slight increase in the N($^4$S) density over the short time-scale considered in figure 14, which is in agreement with TALIF measurements (Figure 11).

This finding leads to a slightly different interpretation from the one given in reference [17] where the observed increase in N($^4$S)-atom density during the early post-dicharge phase of a pulsed plasma was attributed to super-elastic collisions on N($^2$P/$^2$D). Here, we showed that this increase is most probably due to the wall-de-excitation of the metastable state N($^2$D/$^2$P) atoms that results in a net production of N($^4$S), and therefore an increased N($^4$S) density, during 200 µs, which is consistent with the results obtained from TALIF measurements. Note that this short time-scale kinetic effect may be more or less important depending on the discharge conditions. It seemed in particular much more pronounced under the experimental conditions considered in reference [17]. In any case, the contribution of N($^2$D) ad N($^2$P) metastable states to N($^4$S) kinetics should be considered if the N-atom density in a strongly emissive plasma is to be extrapolated from post-discharge density TALIF measurements. Before closing this section, it is worthy to emphasize that the present analysis shows that the assumption of diffusion/recombination dominated by N($^4$S) atom kinetics is only valid when N($^2$P) and N($^2$D) metastable states fully de-excite.

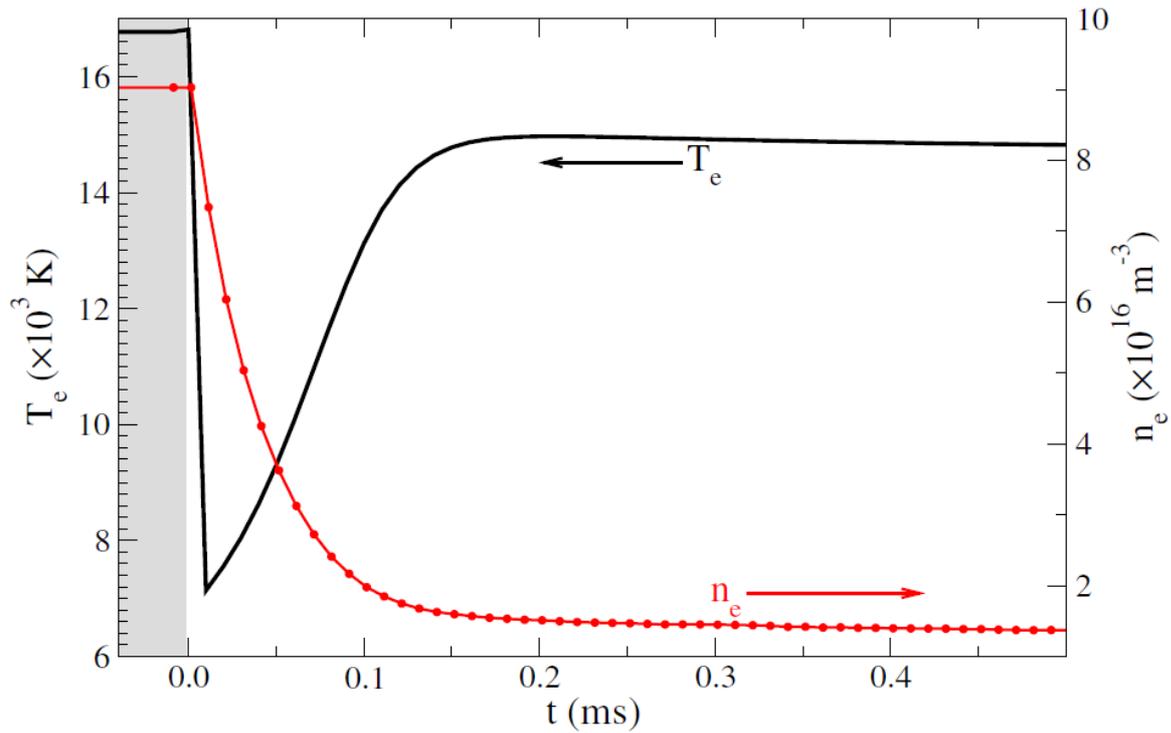

Figure 12 : Time-variation of the electron temperature and electron density. The simulation was performed for a pressure of 20 Pa, a power density of 1 W.cm$^{-3}$, a pulse period of 100 ms and a duty cycle of 0.5. The emphasis is put on the high/low power transition.

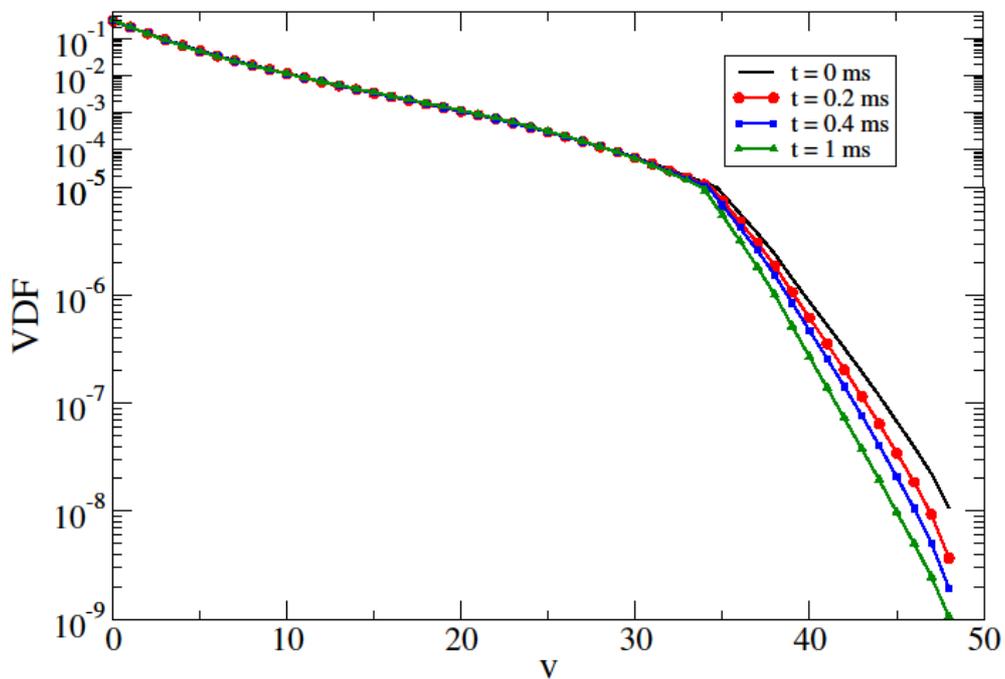

Figure 13 : Time-variation of the vibrational distribution functions (VDF) during the transition to the low power phase. The simulation was performed for a pressure of 20 Pa, a power density of 1 W.cm$^{-3}$, a pulse period of 100 ms and a duty cycle of 0.5. The emphasis is put on the high/low power transition.

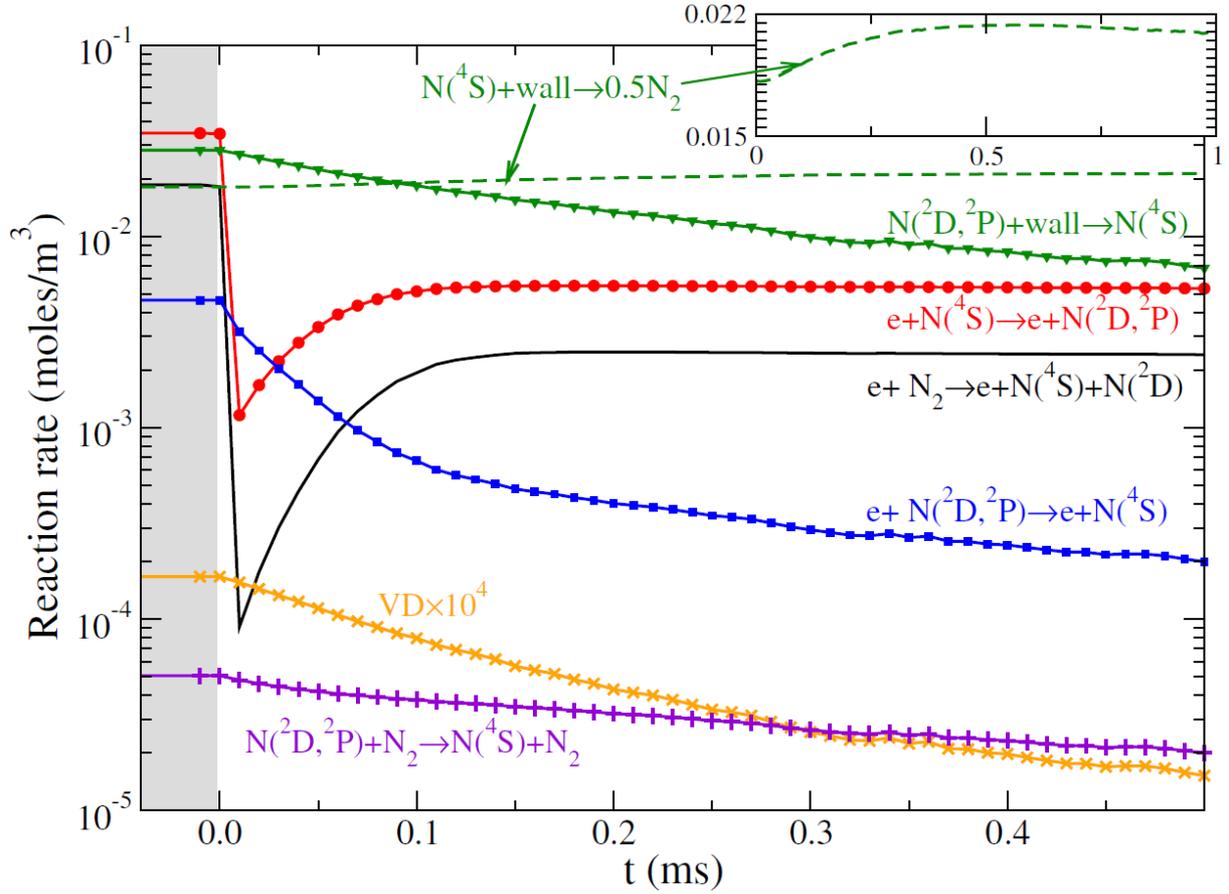

Figure 14: Time-variation of the main electron impact processes involved in the kinetics of N($^4$S)-atom at the transition between the high- and low-power phases. The rate of surface recombination processes are also shown for comparison. Also shown in the insert is the increase of surface recombination of N($^4$S) in the low power phase.

## 5. Conclusion

In this paper we showed that a simplified and straightforward methodology may be used to measure N($^4$S)-atom densities by means of ns-TALIF technique in strongly emissive plasmas. This method makes use of the TALIF intensities measured with the laser central frequency $\nu_L$ tuned to the peak of absorption ($\nu_L = \nu_N/2$), so-called the peak excitation method (PEM), instead of the frequency integration of the laser excitation spectrum (the full excitation method, FEM). It requires however the determination of the statistical overlap factor at $\nu_L = \nu_N/2$. This is performed by a single FEM measurement. The use of this PEM method along with a fitting/filtering procedure developed based on physical considerations enables to strongly reduce the experimental uncertainty. This allows performing measurements even in strongly emitting discharge and/or for pulsed plasmas that could not be investigated using the widely used FEM approach.

The straightforward and simplified methodology is valid as long as the spectral profiles of the absorption lines chosen for TALIF measurements do not change over the investigated plasma conditions. The approach seems to be particularly suitable for low temperature moderate density plasmas where the change in line broadening width due to Doppler and Stark effects remain fairly limited. The developed simple PEM was validated by comparing the resulting density values with those measured using the well-established FEM.

TALIF measurements using the PEM were then used to investigate plasma kinetics in strongly emissive pulsed ECR microwave plasmas. Measurements showed a linear increase in both the N-atom density and the FPS emission intensity with the coupled power, which would indicate a linear increase in the power density with the power (which means that the plasma volume is constant). This consistency between the variations of the power density, the FPS emission and the measured N($^4$S)-atom density further supports the validity of this simplified approach.

The simplified ns-TALIF PEM was also used to investigate the transients at the transition between the low and high power phases of pulsed plasmas. We confirmed the existence of the unexpected increase in N($^4$S)-atom density at the early stage of the low-power phase reported in reference [17]. We used a quasi-homogenous plasma model that involves detailed collisional-radiative models for $N_2$ and N to identify the kinetic effects that lead to the observed density increase. We especially showed that this increase is due to the surface-deexcitation of N($^2$D) and N($^2$P), the role of super-elastic collisions suggested by [17] being much less important in our conditions. From the practical point of view, the results obtained show that such increase should be taken into account when using extrapolated values from N-atom densities measured in the post discharge in order to determine the dissociation yield in an emissive discharge.

The simplified TALIF method discussed in this paper is suitable for a detailed investigation of the dissociation kinetics in a pulsed plasma with strongly emissive power-on phases. It should in particular allow a systematic investigation of the change in the dissociation yield with the pulse period and duty cycle so as to investigate if the $N_2$ dissociation yield may be enhanced in these discharges.

**Acknowledgments**

The authors thank Dr. S Iséni, Dr C. Drag and Dr N. Sadeghi for helpful discussion.

This work was supported by the PRC program of French National Research Agency under the ASPEN project agreement ANR-16-CE30-0004, by the SESAME research and innovation programme of the Ile-de-France Region under the project grant DIAGPLAS and the CNRS "Réseau Plasmas Froids". One of the authors (EB) was supported by the European Union's Horizon 2020 research and innovation programme under the Marie Skłodowska-Curie grant agreement No 665850.